\documentclass[aps,prl,tightenlines,onecolumn,groupedaddress,superscriptaddress,showpacs,nofootinbib]{revtex4}
\usepackage{amssymb}
\usepackage{color,graphicx}
\usepackage{lmodern}
\usepackage{amsmath}
\usepackage{amsbsy}
\usepackage{amsthm}
\usepackage{float}
\usepackage{bbm}
\usepackage{bm}
\usepackage{graphicx}

\newcommand\env{\operatorname{env}}

\begin{document}
\title{ Langevin Equation for a Dissipative Macroscopic Quantum System:\\
Bohmian Theory versus Quantum Mechanics}

\author{Hamid Reza Naeij}
\email[]{naeij@alum.sharif.edu}
\affiliation{Research Group on Foundations of Quantum Theory and Information,
Department of Chemistry, Sharif University of Technology
P.O.Box 11365-9516, Tehran, Iran}
\author{Afshin Shafiee}
\email[Corresponding Author:~]{shafiee@sharif.edu}
\affiliation{Research Group on Foundations of Quantum Theory and Information,
Department of Chemistry, Sharif University of Technology
P.O.Box 11365-9516, Tehran, Iran}

\affiliation{School of Physics, Institute for Research in Fundamental Sciences (IPM), P.O.Box 19395-5531, Tehran, Iran}

\begin{abstract}
In this study, we solve analytically the Schrodinger equation for a macroscopic quantum oscillator as a central system coupled to a large number of environmental micro-oscillating particles. Then, the Langevin equation is obtained for the system using two approaches: Quantum Mechanics and Bohmian Theory. Our results show that the predictions of the two theories are inherently different in real conditions. Nevertheless, the Langevin equation obtained by Bohmian approach could be reduced to the quantum one, when the vibrational frequency of the central system is high enough compared to the frequency of the environmental particles.

\end{abstract}
\pacs{03.65.Ta, 03.65.Yz, 05.10.Gg}
\maketitle
\textbf{keywords} Langevin equation, Macroscopic quantum system, Harmonic environment, Bohmian theory

\section{Introduction}

In recent years, dissipative quantum systems have been studied broadly as a new research field in physics. The study of open quantum systems has recieved great attention due to the importance of environmental effects in description of seminal phenomena such as quantum decoherence, quantum information, quantum tunneling in physical and biological systems, etc. \cite{Weiss,Breuer}. 

Moreover, the investigation of dissipative behavior of macroscopic quantum systems (MQSs), pioneered by Caldeira and Leggett \cite{Caldeira 2}, is an attractive topic in the quantum physics. MQSs are usually considered as a bridge between the quantum and the classical systems and show quantum behavior at the macroscopic scale, rather than at the atomic scale where quantum effects are prevalent. It is commonly believed that $\it{macroscopic}$, here, refers to those situations where a large number of particles are involved. More precisely, it holds when dynamical degrees of freedom for the system is large. The best known examples of MQSs are superconductors and superfluids \cite{leggett0, Garg,leggett2, Caldeira1, Taka}.  

A dissipative quantum system, independent of its macro or micro aspect, is often considered as a system coupled to a large number of harmonic oscillators, modeled as a heat bath. The entire system which consists of one particle plus a thermal reservoir allows us to study the origin of irreversibility in a dynamic approach. In this connection, the common representation of the energy fluctuations and the dissipative effects caused by interaction of the system with its environment is analytically discussed by the quantum Langevin equation \cite{Dorofeyev}.

In a general way, the quantum Langevin equation is nothing but the Heisenberg equation of motion for a macroscopic quantum particle interacting with the environment. The environment could be considered to exert a fluctuating force on the system. The force depends on the motion of the system and could be experienced by the system at rest even for small displacements. Eliminating the environment's variables in the equation of motion of the central system, one reaches a reduced equation which, besides the system's degrees of freedom, involves additional variables describing the fluctuation effects of the environment, usually known as a noise function. This shows that the system exchanges energy with its surrounding. In other words, it is the signature of dissipation \cite{Bhattacharya, Jaekel}.

Many works have been done to describe the dissipative quantum systems by using the quantum Langevin equation in a coherent way. In 1965, Mori derived a generalized Langevin equation for the quantum mechanical operators from the Heisenberg equation of motion, using a projection operator defined in Liouville space \cite{Mori}. Kostin also derived the quantum Langevin equation for a Brownian particle interacting with a thermal bath \cite{Kostin}. In this approach, a dissipative as well as a random potential are both together incorporated in the time-dependent Schrodinger equation, responsible for the thermal and the statistical influences of the environment, respectively. In the particular case of the quantum Langevin equation, Caldeira and Leggett used the path integral method to study the dissipative quantum tunneling \cite{Caldeira 3}. In this approach, a quantum noise function is introduced to show the environmental thermal fluctuations by which the system is affected. This equation can be reformulated in a Hamiltonian-based model, in which the noise arises from a collection of harmonic oscillators coupled to the system. Similarly, in a series of publications, Ford and others obtained various forms of the quantum Langevin equation for a quantum particle coupled to a heat bath \cite{Ford 1, Ford 2, Ford 3}.

Along these efforts, however, Bohmian mechanics has not been widely used yet, within the studies of the dissipative quantum systems. In this regard, the first study was carried out about 22 years ago by Vandyck, who investigated the decay of the harmonic oscillator eigenstates, lossing the energy \cite {Vandyck}. Also, Tilbi and others used Bohmian mechanics to derive an expression for the damping harmonic oscillator, albeit in the Feynman path integral approach \cite{Tilbi}.

What we propose here is to investigate the dissipative behavior of a MQS in interaction with its surrounding. We use a collection of quantum harmonic oscillators as a typical model for the environment. Then, the Langevin equation will be calculated for the central system in both Quantum Mechanics (QM) and Bohmian Theory (BT). We compare the Bohmian damping equation with what is derived by QM, afterwards. Our results show that the predictions of the two competing theories are inherently different in real circumstances. Nevertheless, we show that the Langevin equation obtained in the Bohmian approach can reproduce the quantum one for too large frequencies of the system compared to the environmental ones.

The paper is organized as follows. In section 2, the bilinear harmonic model of the environment is introduced and analyzed. Then, the ground state of the system coupled to $N-$particles of the environment is exactly evaluated. In section 3, the Langevin equation is obtained in QM and BT for a dissipative MQS, respectively. Afterwards, we compare and analyze the ultimate damping equations obtained by these two formalisms and discuss about them in section 4.

\section{Harmonic model of the environment as a typical form of the open quantum system }

First, let us begin with the way one can introduce dimensionless parameters for an arbitrary quantum sysytem \cite{Taka}. To do this, we define the characteristic parameters for length $R_0$ and energy $U_0$ as constant units of length and energy, respectively. Subsequently, for a particle of mass $M$, one can define the characteristic time as $\tau_0=R_0 / (U_0/ M)^\frac{1}{2}$. Since, $U_0$ acts like the kinetic energy of a given system, the unit of momentum could be defined as  $P_0=(U_0 M)^\frac{1}{2}$. Then, the conjugate variables $q$ (position) and $p$ (momentum) are defined as $q=R/R_0$ and  $p=P/P_0$, respectively.

Here, we assume that the central system is a quantum harmonic oscillator subjected to a collection of micro-oscillators in the environment. The total Hamiltonian could be written as the sum of the Hamiltonian of the system $H_s$, the environment $H_e$ and the interaction $H_{se}$:
\begin{equation}
\label{eq1}
H=H_s+H_e+H_{se}
\end{equation}

In the dimensionless form, the Hamiltonians in (1) could be defined as:
\begin{equation}
\label{eq2}
H_s=\frac{p^2}{2}+V({q})
\end{equation}
\begin{equation}
\label{eq3}
H_e={\sum _{\alpha=1}^{N} }\Big[ \frac{p^2_\alpha}{2}+\frac{\omega^2_\alpha}{2}{x^2_\alpha-}\frac{\bar h\omega_\alpha}{2}\Big]
\end{equation}
\begin{equation}
\label{eq4}
 H_{se}=-\sum _{\alpha=1}^{N} {\omega}^2_\alpha f_\alpha(q) x_\alpha+\frac{1}{2}\sum _{\alpha=1}^{N} \omega^2_\alpha \big( f_\alpha (q)\big) ^2
\end{equation}

\noindent where $N$ is the total number of the environmental particles and $\omega_\alpha$ is the frequency of the particle $\alpha$ in the environment. Also, $V(q)$ denotes the potential of the system and $f_\alpha (q)$ is an arbitrary function of the space variable of the system $q$, coupled to the position $x_\alpha$ of the environmental particles $\alpha$.

The canonical relations are $[\hat q,\hat p]=i\bar h$ and $[\hat x_\alpha, \hat p_\beta]={i\bar h}\delta_{\alpha\beta}$, where
 \begin{equation}
 \label{eq5}
 \bar h=\frac{\hbar}{{P_0}{R_0}}
 \end{equation}
  As one can see in (3) and (5), instead of Planck constant $\hbar$, a new dimensionless parameter $\bar h$ appears which quantitavely rates the quantum nature of the system. Strictly speaking, the situation in which one gets $\bar h\ll1$, the system behaves quasi-classically \cite{Taka}. In many applications, $R_0$ is defined as characteristic length of resonance between left and right counterparts of a double-well potential. So, $\bar h$ in (5) could be also written as:
\begin{equation}
\label{eq6}
\bar h=\lambdabar_0/R_0
\end{equation}
\noindent where $\lambdabar_0=\lambda_0 / 2\pi$. Here, $\lambda_0$ denotes the characteristic wavelength of the central system. For a MQS, $\lambda_0$ is too small compared to $R_0$ which is nearly a fixed value for known models of potential. This is a demonstration of how the condition $\bar h\ll1$ can display the classicality of the system. Notice that the macroscopic feature of the system depends on its physical properties such as size and wavelength which leads to $\bar h\ll1$ for quasi-classical situation. (For more details, see \cite{Naeij1, Naeij2}).

Hereafter, for simplicity, we assume that $ f_\alpha(q)={\gamma_\alpha}q$ in (4) where $ 0<\gamma_\alpha<1$ denotes the strength of the coupling between the system and the environment. Using the above assumption, one reaches the conclusion that $H_{se}$ is linear both to $x_\alpha$ and $q$. So, the model is called $\it{bilinear}$. Regarding the bilinear assumption, the total form of the Hamiltonian can be written as:
\begin{align}
\label{eq7}
H=\frac{p^2}{2}+V_1(q)+\sum _{\alpha=1}^{N} \Big[\frac{p^2_\alpha}{2}+\frac{\omega^2_\alpha}{2}x^2_\alpha(1-\gamma_\alpha)-\frac{\bar h\omega_\alpha}{2}\Big] +\frac{1}{2}\sum _{\alpha=1}^{N}  \gamma_\alpha\omega^2_\alpha(x_\alpha-q)^2
\end{align}
where $V_1(q)=V(q)-\frac{1}{2}\big( \sum_\alpha{\gamma_\alpha}(1-\gamma_\alpha)\omega^2_\alpha\big) q^2$.

\noindent Here, $V(q)=\frac{1}{2}\omega_0 ^2 q^2$ and $ \omega_0 $ are the potential and the vibrational frequency of the central system, respectively. In (7), one sees that the particle feels the efficient potential $V_1(q)$. Moreover, the main coupling is between each environmental oscillator $\alpha$ with spring constant $(1-\gamma_\alpha)\omega^2_\alpha$ and the central particle having the spring constant $\gamma_\alpha\omega^2_\alpha$. In the next part, we obtain the stationary eigenfunctions of the total Hamiltonian for the ground state of the entire system.

\subsection{Analytical solution of Schrodinger equation}
Considering the bilinear assumption, the total Hamiltonian in (7) can be represented as:
\begin{equation}
\label{eq8}
H=\frac{p^2}{2}+\frac{1}{2}\omega^2 q^2+\sum _{\alpha=1}^{N} \Big[ \frac{p^2_\alpha}{2}+\frac{\omega^2_\alpha}{2}x^2_\alpha-\omega^2_\alpha\gamma_\alpha q x_\alpha\Big]
\end{equation}
where $\omega=[\omega_0 ^2+\sum_\alpha \omega^2_\alpha \gamma^2_\alpha]^\frac{1}{2}$.

Here, we define:
\begin{equation}
\label{eq9}
H_\alpha=\frac{p'^2}{2}+\frac{1}{2}\omega^2 q'^2+ \frac{p^2_\alpha}{2}+\frac{1}{2} \omega^2_\alpha x
^2_\alpha-\omega^2_\alpha \gamma_\alpha q' x_\alpha
\end{equation}
where $p'=p/N , q'=q/N$. The Hamiltonian (8) now reads as the sum of the individual Hamiltonians indexing $\alpha$.

Now, we decouple the Hamiltonian by using the rotation of the position coordinates $(q',x_\alpha)$ and the momentums $(p',p_\alpha)$ to define new position and momentum coordinates, $(x_+,x_-)$ and $(p_+, p_-)$, respectively:
\begin{equation}
\label{eq10}
\begin{pmatrix}

x_+  \\
x_-
\end{pmatrix}
\begin{matrix}\\\mbox{}\end{matrix}
=\begin{pmatrix} \cos\theta & \sin\theta \\ -\sin\theta & \cos\theta \end{pmatrix}
\begin{pmatrix} q' \\
x_\alpha
\end{pmatrix}
\end{equation}
\noindent and
\begin{equation}
\label{eq11}
\begin{pmatrix}

p_+  \\
p_-
\end{pmatrix}
\begin{matrix}\\\mbox{}\end{matrix}
=\begin{pmatrix} \cos\theta & \sin\theta \\ -\sin\theta & \cos\theta \end{pmatrix}
\begin{pmatrix} {p'} \\
{p}_\alpha
\end{pmatrix}
\end{equation}

\noindent Under the rotation, the kinetic energy parts of the Hamiltonian (9) remains invariant. Thus, decoupling of the Hamiltonian is achieved by diagonalizing the potential energy. Defining

\begin{equation}
\label{eq12}
\theta=\frac{1}{2}\arctan\Big[ \frac{\omega'^2_\alpha}{\omega^2 -\omega^2_\alpha}\Big]
\end{equation}
under the rotation, the Hamiltonian is transformed to
\begin{equation}
\label{eq13}
H_\alpha=\frac{p_{+\alpha}^2}{2}+\frac{1}{2} \omega_{+\alpha}^2 x_{+\alpha}^2+\frac{p_{-\alpha}^2}{2}+\frac{1}{2} \omega^2_{-\alpha} x_{-\alpha}^2
\end{equation}
where $H_\alpha =H_{+\alpha}+H_{-\alpha}$. Denoting $\omega'_\alpha=i ( 2{\omega^2_\alpha}{\gamma_\alpha})^\frac{1}{2}$, we have

\begin{equation}
\label{eq14}
\omega_{+\alpha}=\big[ \omega^2 \cos^2\theta+\omega^2_\alpha \sin^2\theta+\omega'^2_\alpha \sin\theta\cos\theta\big]^\frac{1}{2}
\end{equation}
\begin{equation}
\label{eq15}
\omega_{-\alpha}=\big[\omega^2 \sin^2\theta+\omega^2_\alpha \cos^2\theta-\omega'^2_\alpha \sin\theta\cos\theta\big]^\frac{1}{2}
\end{equation}

Considering the definitions $a=\sin\theta$, $b=\cos\theta$ and $c=\tan2\theta <0$ (i.e., $\omega^2>\omega_\alpha^2$), the plus-minus position coordinates of the particles of the environment can be written as:
\begin{align}
\label{eq16}
x_{+\alpha}&=bq'+a{x}_\alpha \nonumber\\
x_{-\alpha}&=-aq'+b{x}_\alpha
\end{align}
\noindent where $\alpha=1$ to $N$. So, the wave function of the ground state of the Hamiltonian (8) is obtained as \cite{Naeij1, Naeij2}
\begin{equation}
\label{eq 17}
\psi_0(q, x_1,..., x_N)=\prod_{\alpha=1}^{N}\Big(\frac{{\omega_{+\alpha}} {\omega_{-\alpha}}}{{\pi^2\bar h^2}}\Big)^\frac{1}{4}\exp\Big(\frac{-{\omega_{+\alpha}}{x_{+\alpha}^2}}{2\bar h}\Big)\exp\Big(\frac{-{\omega_{-\alpha}}{x_{-\alpha}^2}}{2\bar h}\Big)
\end{equation}

\section{Langevin Equation for a Dissipative Macroscopic Quantum System}
\subsection{Quantum Approach}

The quantum Langevin equation is derived by considering the Heisenberg equation of motion for the central macro-system which interacts with the harmonic environment in a linear fashion \cite{Taka}. In the Heisenberg picture, by using the Hamiltonian (7), $q(t)$ obeys the following equation of motion:

\begin{equation}
\label{eq18}
\frac{d^2 q(t)}{dt^2}=- \tilde{V'} \big (q(t)\big)+\sum _{\alpha=1}^{N} \omega^2_\alpha \big [ f'_\alpha \big(q (t)\big)  x_\alpha \big]
\end{equation}
Here,
\begin{equation}
\label{eq19}
\tilde{V} ( q(t))=V(q)+\frac{1}{2}\sum _{\alpha=1}^{N} \omega^2_\alpha \big( f_\alpha (q)\big) ^2
\end{equation}
where the prime $'$ in $\tilde{V'} \big(q(t)\big)$ and $f'_\alpha \big(q (t)\big)$ denotes $q-$differentiation. Now, we define Ohmic distribution for the frequencies of the environmental harmonic oscillators $\omega_e$ as
\begin{equation}
\label{eq20}
J(\omega_e) \simeq \lambda'_Q \omega_e e^{-\omega_e/\omega_c}
\end{equation}
where $\lambda'_Q$ is a positive constant which quantifies the strength of the interaction between the environment and the macro-system and $\omega_c$ is the cut-off frequency which maximizes $J(\omega_e)$.

By using the bilinear model, after some calculations, one can derive the equation of motion of the central system as

\begin{equation}
\label{eq21}
\frac{d^2 q(t)}{dt^2}+{V'} \big(q(t)\big)+\int_{t_0}^{t}T(t-t') \frac{dq(t')}{dt'} dt'= R(t)
\end{equation}
 where $T(t)=\sum_\alpha {\bar \gamma_\alpha}^2\omega_\alpha^{-1} \cos {\omega_\alpha t}, $ $(\bar \gamma_\alpha=\gamma_\alpha{\omega_\alpha}^{3/2})$ is called the retarded resistance function \cite{Taka}. This function could be expressed in terms of the environmental frequency distribution $J(\omega_e)$ as
 \begin{equation}
 \label{eq22}
 T(t)=\frac{2}{\pi} \int_{0}^{\infty} \frac{d \omega_e}{\omega_e} J(\omega_e) \cos \omega_e t
 \end{equation}
 where $J(\omega_e)=\frac{\pi}{2}\sum_{\alpha}{\bar \gamma_\alpha}^2 \delta(\omega_e-\omega_\alpha)$, defined as a general relation.
 
 For an Ohmic distribution (20), one can show that 
 
 \begin{equation}
 \label{eq23}
 T(t-t_0)\approx \frac{2\lambda'_Q \omega_c}{\pi}\big(\frac{1}{1+\omega^2_c (t-t_0)^2}\big)
 \end{equation}
 
 We also define:
 \begin{equation}
 \label{eq24}
 R(t)=\sqrt{\frac{\bar h}{2}}\sum _{\alpha=1}^{N} {\bar \gamma_\alpha} \big[ e^{-i(t-t_0)\omega_\alpha }({b_\alpha}-\sqrt \frac{\omega_\alpha}{2\bar h}\gamma_\alpha q)+h.c \big]
 \end{equation}
where {\it h.c} means Hermitian conjugate, $ b_\alpha$ and $ b_\alpha^\dagger$ are the creation and the annihilation operators of the environmental particles, respectively.

An equation of the form (21) is called the {\it quantum Langevin equation}. The third term in the left-hand side of (21) implies the existence of the frictional resistance retarded in time. The right-hand side may be interpreted as representing a fluctuating force of the environment on the macro-system, known as a noise term.

We consider the relation (17) as the initial state of the entire system (the macro-system plus the environment). Since $R(t)$ in (24) depends on the environmental variables $x_\alpha$ and $p_\alpha$ (via $b_\alpha$ and $ b_\alpha^\dagger$), by averaging over such variables, one gets:
\begin{equation}
\label{eq25}
\ddot q(t) +V'\big(q(t)\big)+M(t)=\langle R(t) \rangle_{\env}
\end{equation}
where 
\begin{equation}
\label{eq26}
{M}(t)=\int_{t_0}^{t}T(t-t') {\dot q}(t') dt'=-\int_{t-t_0}^{0}T(t'){\dot q}(t+t') dt'
\end{equation}
is the memory term. For the temporal domain $1\ll \omega_c (t-t_0) $ and using Markov approximation, the memory term for Ohmic frequency distribution \big(defined in (20)\big) becomes $M (t)\simeq  \lambda'_Q {\dot q}(t)$ \cite{Taka}. In other words, the nature of the dissipation is contained in the memory function. 

The Markov assumption have means that ${\dot q}(t+t')\approx {\dot q}(t)$. This is because, regarding the wave function (17), one can show that the momentum probability distribution of the central system is Gaussian-type \cite{Naeij2}, which remains stationary at time. If the potential terms of the central system (including the $q$ variable) would be neglected compared to $p^2/2$ in (8), such assumption is valid, specially for MQSs with weak interactions.

One can also write $ R(t)$ as 

\begin{equation}
\label{eq27}
R(t)\simeq\sqrt{\frac{\bar h}{2}}\sum _{\alpha=1}^{N} {\bar \gamma_\alpha} \big[ e^{-i(t-t_0)\omega_\alpha }{b_\alpha}+h.c \big]
\end{equation}

The term neglected in (24) is of the type 

\begin{equation}
\label{eq28}
-q\sum_\alpha \gamma^2_\alpha \omega^2_\alpha \cos[\omega_\alpha(t-t_0)]=-q T(t-t_0)
\end{equation}
where $T(t-t_0)$ was defined in (22). For $\omega_c (t-t_0)\gg 1$ and considering $J(\omega_\alpha)$ in (20), the retarded resistance function $T(t-t_0)$ approaches zero, as is obvious in (23).

Moreover, for the state function $\psi_0(q, x_1,..., x_N)$ in (17) at $t=0$, $\langle x_{\pm \alpha}\rangle =0 $, as well as $\langle p_{\pm \alpha}\rangle =0 $. Hence, from (16), one concludes that $\langle x_ \alpha\rangle =0 $, similarly $\langle p_ \alpha\rangle =0 $. As a result, $\langle{R}(t) \rangle_{env}\simeq0$ in (27), because $b_\alpha$ and $ b_\alpha^\dagger$ are defined in terms of $ x_\alpha$ and $ p_\alpha$. Consequently, the quantum Langevin equation in (25) becomes
\begin{equation}
\label{eq29}
{\ddot q}(t)+2\omega_0\lambda_Q {\dot q}(t)+V'\big(q(t)\big)\simeq 0
\end{equation}
where $\lambda'_Q= 2\omega_0 \lambda_Q$ and $\lambda_Q$ is the usual friction coefficient. Also, $V'\big(q(t)\big)=\frac{\partial}{\partial q}(\frac{1}{2}\omega_0^2 q^2)$ is the force applied on the central system at $t>0$. For a MQS having a nearly sharp distribution of position $q$ (see, e.g., \cite{Naeij1}), the final form of the quantum langevin equation becomes
\begin{equation}
\label{eq30}
\ddot q_Q(t)+2\omega_0\lambda_Q \dot q_Q(t)+\omega_0^2 q_Q(t)\simeq 0
\end{equation}
where the notation $q_Q$ instead of $q$ is used to show the quantum approach.

\subsection{Bohmian Approach}

Let us express the wave function of the central system (with space variable $q$) plus the $N$-particles of the environment (with space coordinates $x_1, x_2, ..., x_N$) in Bohmian approach as the following relation:
\begin{equation}
\label{eq31}
\psi(q, x_1,..., x_N ,t)=R \exp (iS/\bar h)
\end{equation}
where $R=R(q, x_1,..., x_N, t)$ and $S=S(q, x_1,..., x_N, t)$ are real functions of space coordinates and time. The function $S$ is a sort of action and is measured in unit of $\bar h$ and $R$ the real amplitude of probability density $\vert \psi\vert^2$. Inserting (31) into the time-dependent Schrodinger equation and separating it in real and imaginary parts, one obtains the following eqautions in the dimensionless form:

\begin{equation}
\label{eq32}
\frac{\partial S}{\partial t}+\frac{1}{2}(\frac{\partial S}{\partial q})^2- \frac{\bar h^2}{2}\frac{1}{R}\frac{\partial^2 R}{\partial q^2}+\sum _{\alpha=1}^{N} \frac{(\overrightarrow\nabla_\alpha S)^2}{2}-\frac{\bar h^2}{2}\sum _{\alpha=1}^{N}\frac{\hat \nabla_\alpha ^2 R}{R}+V=0
\end{equation}
and
\begin{equation}
\label{eq33}
\frac{\partial R^2}{\partial t}+\frac{\partial}{\partial q}(R^2\frac{\partial S}{\partial q})+\sum _{\alpha=1}^{N} \overrightarrow\nabla_\alpha .{(R^2\overrightarrow\nabla_\alpha S)}=0
\end{equation}

\noindent where $V$ is the classical potential energy which includes interparticle and external potentials. The relation (32) is similar to the  classical Hamilton-Jacobi equation, apart from the extra term

\begin{equation}
\label{eq34}
Q(q, x_1,..., x_N, t)=- \frac{\bar h^2}{2}\frac{1}{R}\frac{\partial^2 R}{\partial q^2} -\frac{\bar h^2}{2}\sum _{\alpha=1}^{N}\frac{\hat\nabla_\alpha ^2 R}{R}
\end{equation}

The function $Q$ is called {\it quantum potential}. Moreover, we have:

\begin{equation}
\label{eq35}
\ddot{q}=-\frac{\partial}{\partial q} (Q+V)
\end{equation}

The relation (35) has the form of Newton's second law, in which the central system is subjected to a quantum force $-\partial Q/\partial q$ in addition to the classical force $-\partial V/ \partial q$ \cite{Bohm 1, Bohm 2, Holland}.

If we calculate the quantum potential for the central system coupled to $N$-environmental particles, using the wave function obtained in (17), one gets
\begin{align}
\label{eq36}
Q\big(q(t), x_1(t) ,..., x_N(t)\big)=&\frac{\bar h}{4}\Big \lbrace \big [2A-\frac{1}{2\bar h}(4A^2q^2+4ABq+B^2)\big ]
\nonumber\\
+&\sum _{\alpha=1}^{N}\big [(2a^2\omega_{+\alpha}+2b^2\omega_{-\alpha})-\frac{1}{2\bar h}\big(\frac{2abq}{N}(\omega_{+\alpha}-\omega_{-\alpha})+2a^2\omega_{+\alpha}x_\alpha+2b^2\omega_{-\alpha}x_\alpha\big)^2\big ]\Big \rbrace
\end{align}
where 
\begin{equation}
\label{eq37}
A=\sum _{\alpha=1}^{N} \frac{1}{N^2} [a^2\omega_{-\alpha}+b^2 \omega_{+\alpha}]
\end{equation}
and
\begin{equation}
\label{eq38}
B=2ab\sum _{\alpha=1}^{N}\frac{1}{N}\big[x_\alpha (\omega_{+\alpha}-\omega_{-\alpha})\big]
\end{equation}

As $\bar h\rightarrow 0$, which illustrates the limit at which the macroscopic aspect of the system becomes significant, $Q$ does not vanish but tends to independent terms of $\bar h$. So, the quantum feauture of the central system remains, even in new macroscopic conditions.

The classical potential, as included in (8), is:
\begin{equation}
\label{eq39}
V\big(q(t), x_1(t),..., x_N(t)\big)=\frac{1}{2}\omega^2q^2+\sum _{\alpha=1}^{N}\Big[\frac{\omega^2_\alpha}{2}x^2_\alpha - \gamma_\alpha q \omega^2_\alpha x_\alpha\Big] 
\end{equation}

Using (35) and introducing the term responsible for friction as a (constant)$\times \dot q$, the equation of motion of the central system could be written as
\begin{equation}
\label{eq40}
\ddot q_B+ 2\xi \lambda_B \dot q_B+\xi^2 q_B = \eta \big(x_1(t),..., x_N(t)\big)
\end{equation}
where $q_B=q_B (x_1,..., x_N,t)$ and $\lambda_B$ is the friction constant in Bohmian approach. This equation has the form of the Langevin equation. Hence, we call it the {\it Bohmian Langevin equation}. In (40), we have
\begin{equation}
\label{eq41}
\xi^2=-A^2+\omega^2-a^2b^2 \sum _{\alpha=1}^{N}\frac{1}{N^2}(\omega_{+\alpha}-\omega_{-\alpha})^2
\end{equation}
\noindent and
\begin{equation}
\label{eq42}
\eta \big(x_1(t),..., x_N(t)\big)= \frac{AB}{2}+\sum _{\alpha=1}^{N}(\gamma_\alpha  \omega^2_\alpha x_\alpha) +\sum _{\alpha=1}^{N}\frac{1}{N}\big [ab(\omega_{+\alpha}-\omega_{-\alpha})\big]\big[a^2\omega_{+\alpha}x_\alpha+b^2\omega_{-\alpha}x_\alpha\big)\big ]
\end{equation}

By averaging over the spatial coordinates of the environmental particles in (40) \big(using $\vert \psi_0\vert^2=\vert R\vert^2$, where $\psi_0$ is defined in (17)\big), the equation of motion of the central system becomes:
\begin{equation}
\label{eq43}
\ddot q_B(t)+ 2\xi \lambda_B \dot q_B(t)+\xi^2 q_B(t) = 0
\end{equation}
\noindent which has a similar form of the quantum Langevin equation in (30). Here, we have $\langle \eta \big(x_1(t),..., x_N(t)\big)\rangle=0$, since $\langle x_\alpha\rangle=0$ in all the terms of the relation (42). 

It is also important to note that the time evolution considered here is also Markovian-type. Since, just like before, the same wave function (17) is used in Bohmian dynamics with the same resulted momentum and position distributions. For strong interaction and/or dominant potential terms in (8), the Markov approximation is not valid and the relations (30) and (43) should be improved.  

\section{Results and Discussion}

We obtained the Langevin equation for a macro-system coupled to $N$-environmnetal micro-oscillating particles in two different approaches: QM and BT. The general form of the equations (30) and (43) is similar, in which the mean value of the noise term or the fluctuation force is zero in both cases. The main differences between the two Langevin equations is related to the frequency of the macro-system in QM ($\omega_0$) and the one derived from BT ($\xi$).

The Langevin equations (30) and (43) leads to the following relations in the Quantum and the Bohmian approaches, respectively
\begin{equation}
\label{eq44}
q_Q(t)=C e^{-\omega_0 \lambda_Q t} \sin (\sqrt{1-\lambda_Q^2} \omega_0 t+\phi)
\end{equation}
\begin{equation}
\label{eq45}
q_B(t)=C' e^{-\xi \lambda_B t} \sin (\sqrt{1-\lambda_B^2} \xi t+\phi')
\end{equation}

In these equations, $C , C'$ are the amplitudes and $\phi, \phi'$ are the phases of motion, which can be related to the initial conditions. As is clear in (44) and (45), the central system shows  damping behavior due to the interaction with the environmental particles. Yet, only in the Bohmian Langevin equation, the environmental parameters like $\gamma_\alpha$ and $\omega_\alpha$ are included via the frequency $\xi$. 

Regarding the relations (37) and (41), one can notice that the second term in the relation (41) is always dominant, since for large values of $N$, the first and the third terms are both diminished. So that $\xi^2\approx\omega^2$, where $\omega^2=\omega_0^2+\sum_{\alpha}\gamma_\alpha^2\omega_\alpha^2$. Then, the equation of motion (45) could be viewed in two different regimes.

In the first situation, which is the case for the most known MQSs, we have $\omega_0\ll\omega_c$ \citep{Taka}. For an Ohmic distribution defined in (20), one reads

\begin{equation}
\label{eq46}
\sum_{\alpha}\gamma_\alpha^2\omega_\alpha^2=\frac{2}{\pi} \int_{0}^{\infty} \frac{J(\omega_e)}{\omega_e} d\omega_e\approx \frac{2}{\pi} \lambda'_Q \omega_c
\end{equation}

\begin{figure}[H]
\centering
\includegraphics[scale=1]{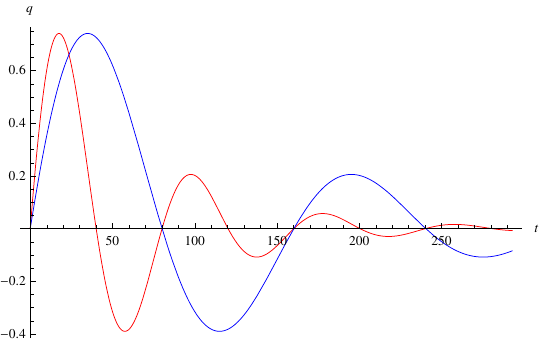}
\caption {The Quantum ({\it blue curve}) and the Bohmian ({\it red curve}) Langevin equations of motion for a macro-system coupled to $N-$environmental particles.}\label{fig1}
\end{figure}

Thus, $\xi^2 \approx \frac{2}{\pi} \lambda'_Q \omega_c$ with $\lambda_Q'=2\omega_0 \lambda_Q$. In other words, when we use Ohmic distribution in Bohmian approach, the relation (45) does not depend on the environmental parameters such as $\omega_\alpha$ and $\gamma_\alpha$, similar to the relation (44). In such regime, however, the difference between the two relations (44) and (45) is significant. For, e.g., $\omega_0 \approx 0.04$ (corresponding to the frequencies of the order of $ 4\times 10^{10}$ $s^{-1}$), $\omega_c \approx 1$ (corresponding to about $10^{12}$ $s^{-1}$)  and $\lambda_Q\approx 0.2$, the quantum and the Bohmian equations of motion are sketched in Fig. \ref{fig1}.

The important point is that both Quantum and Bohmian Langevin equations have been compared in the time domain $1\ll \omega_c (t-t_0)$ in Fig. \ref{fig1} where we used $\omega_c \approx 1$ and $(t-t_0)\approx 300$ (corresponding to about $3\times10^{-10}$ $s$). 

Secondly, one could assume that $\omega^2\approx\omega^2_0$ which is appropriate for micro-systems with $1\ll \omega_c (t-t_0)\ll \omega_0 (t-t_0)$ in the aforementioned time domain. This means that the frequency of the central system is larger than the cut-off frequency of the environment. For such systems, the two equations of motion merge into one equation described by

\begin{equation}
\label{eq47}
q(t)=e^{-\omega_0 \lambda t}\big[\sin (\sqrt{1-\lambda^2} \omega_0 t)(\frac{\dot q(0)}{\sqrt{1-\lambda^2}\omega_0})\big]
\end{equation}
which is another version of (44) with $\lambda_Q=\lambda$ and $q(0)=0$. 

As a result, the more classical trait of the central system is enhanced \big(corresponding to the smaller values of its wavelength in (6)\big), the more discrimination is observed in the Quantum and the Bohmian approaches. For micro-systems with high enough frequencies, however, the effect of the environment is not such that one can observe any difference between the predictions of these two formalisms for describing a quantum-like Langevin equation. This is in accordance with general belief that for microscopic systems, there should be no difference between the predictions of QM and BT.

\end{document}